\documentclass[10pt,conference]{IEEEtran}
\IEEEoverridecommandlockouts

\usepackage{cite}
\usepackage{booktabs}
\usepackage{graphicx}
\usepackage{microtype}
\usepackage{tabularx}
\usepackage{url}
\usepackage{xspace}
\usepackage[hidelinks]{hyperref}
\usepackage{orcidlink}

\newcommand{\fx}{\textsc{f2x50}\xspace}

\begin{document}

\title{f2x50: A Repository-Scale Benchmark for Fortran-to-X Software Migration}

\author{%
\IEEEauthorblockN{Shraddha Surana\textsuperscript{*}\orcidlink{0000-0002-3009-3178}\quad
Ashwin Srinivasan\orcidlink{0000-0002-4911-0038}}
\IEEEauthorblockA{Department of Computer Science and Information Systems,\\
BITS Pilani, K. K. Birla Goa Campus, NH 17B, Zuarinagar, Goa 403726, India\\
\{p20220031, ashwin\}@goa.bits-pilani.ac.in}
\thanks{\textsuperscript{*}Corresponding author: Shraddha Surana.}
}

\maketitle

\begin{abstract}
Repository-level migration requires more than translating isolated functions: it must preserve behaviour across modules, shared state, build structure, and dependencies. Existing translation benchmarks predominantly cover small programs or small repositories, or fixed and widely studied language pairs. We present \fx, a target-independent benchmark comprising 50 public Fortran repositories pinned to exact Git revisions. It contains 12,116 Git-tracked Fortran files and 6,569,806 lines, divided into three structural-complexity tiers and paired with 561 deterministic behavioural probes. Of these probes, 446 compile pinned source files with small drivers and span 131 repository--routine pairs. Every repository has at least five direct checks that collectively execute at least two routines. \fx provides target-independent tools for acquisition, validation, licence review, and reproducibility, together with a reference adapter for evaluating Python targets. The remaining probes are explicitly classified as extractions, inline reimplementations, or synthetic fixtures. The benchmark therefore supports comparable evaluations across migration tools and target languages.
\end{abstract}


\section{Motivation and Contributions}

Function-level translation benchmarks such as CodeXGLUE~\cite{lu2021codexglue} and HumanEval-X~\cite{zheng2023codegeex} omit the repository structure that makes real migration difficult. Recent repository-level resources address part of this gap. RepoTransBench spans 13 language pairs~\cite{wang2026repotransbench}; AlphaTrans evaluates Java-to-Python translation on ten projects~\cite{ibrahimzada2025alphatrans}; FreshBrew targets Java-version upgrades in 228 repositories~\cite{may2026freshbrew}; and RepoMod-Bench provides implementation-agnostic tests for 21 repositories through standardised interfaces~\cite{li2026repomod}. DepWareTrans studies dependency-consistent, incremental migration across co-executable JVM and .NET language pairs~\cite{chand2026depwaretrans}. However, none focuses on scientific Fortran. Its modules, array conventions, global state, numerical tolerances, and varied build practices make it a distinct migration setting. Table~\ref{tab:positioning} summarises these differences.

Fortran remains embedded in long-lived scientific and engineering systems. Their accumulated numerical methods, validation knowledge, and performance tuning make wholesale replacement costly. Migration can connect these systems to modern language ecosystems, development tools, workflows, and computing platforms. Evaluating such migration requires repository-scale artefacts that retain the numerical and system behaviour on which scientific results depend.

\fx addresses this setting without prescribing a target language or migration method. It contributes: (1)~50 real repositories fixed at full 40-character Git revisions, with a revision-specific licence review; (2)~a target-independent catalogue of 561 versioned behavioural probes with fixed inputs and comparison rules; and (3)~reproducible acquisition and validation tools, schemas, and documentation, together with a reference adapter for Python targets. Adapters can use the same repositories and probes to evaluate Python, Rust, Julia, C++, or other target implementations. They must preserve the fixed inputs, comparisons, and tolerances, and report an outcome for all 561 cases.

An earlier development snapshot demonstrates the use of the benchmark in comparative migration studies. Surana et al. reused all 50 pinned repositories and a 382-case version of the behavioural suite to evaluate five Fortran-to-Python migration configurations under identical source-oracle inputs~\cite{surana2026latent}. Holding the repositories, test cases, and comparison rules fixed showed that the primary difference was behavioural coverage i.e. whether a translated repository exposed runnable targets, rather than output accuracy alone. Version~1.0 is the first stable release and this paper its first complete description.

\begin{table*}[t]
\caption{Comparison with representative translation and modernisation benchmarks. Scale is reported as published by each resource; unlike \fx, several resources define a fixed target or task.}
\label{tab:positioning}
\centering
\footnotesize
\setlength{\tabcolsep}{4.2pt}
\begin{tabularx}{\textwidth}{l l l X}
\toprule
Resource & Unit / scale & Languages or task & Behavioural evidence and distinguishing boundary \\
\midrule
CodeXGLUE / HumanEval-X~\cite{lu2021codexglue,zheng2023codegeex} & Functions / problems & Multiple & Task-level inputs and tests; repository structure is absent \\
CodeRosetta~\cite{tehrani2024coderosetta} & 33 paired test programs & Fortran $\leftrightarrow$ C++ & Bidirectional translation; paired test set evaluated primarily with CodeBLEU \\
RepoTransBench~\cite{wang2026repotransbench} & 1,897 repository samples & 13 pairs, 7 languages & Executable repository tests; multilingual pair-specific translation \\
AlphaTrans / FreshBrew~\cite{ibrahimzada2025alphatrans,may2026freshbrew} & 10 / 228 repositories & Java$\rightarrow$Python / Java upgrades & Translated or project tests; fixed target or modernisation task \\
RepoMod-Bench~\cite{li2026repomod} & 21 repositories, 1.6M lines & 8 source languages & Hidden interface tests; implementation-agnostic modernisation \\
\textbf{f2x50} & \textbf{50 repositories, 6.57M lines} & \textbf{Fortran$\rightarrow$X} & \textbf{Revision-pinned repositories with versioned, target-independent behavioural checks} \\
\bottomrule
\end{tabularx}
\end{table*}

\begin{figure}[t]
  \centering
  \includegraphics[width=\columnwidth]{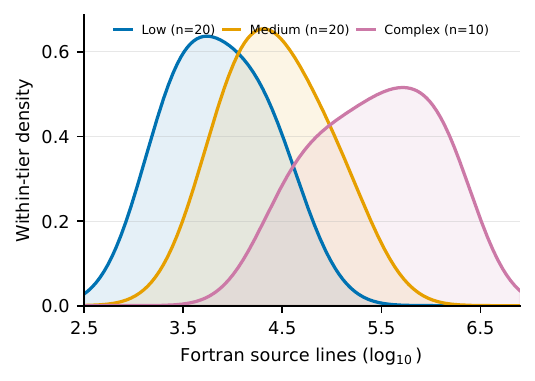}
  \caption{Repository-size distributions in the low, medium, and complex tiers. Gaussian kernel-density estimates use a shared bandwidth over the base-10 logarithm of Fortran source lines. Their overlap shows that size alone does not determine a repository's tier.}
  \label{fig:overview}
\end{figure}

\section{Dataset Construction}

\subsection{Source and selection}

Candidates had to be publicly available, contain at least 1,000 lines of Fortran source, and be practical to compile with a standard Fortran toolchain. We additionally excluded repositories with a known public Python translation at the time of selection, so that the bundled Python reference adapter is not evaluated on cases it could answer by lookup; this exclusion does not apply to future adapters for other target languages. The manifest retains a repository-specific domain label. Table~\ref{tab:domains} groups the repositories into six high-level application areas. These groups distinguish scientific applications from the language, data, and developer infrastructure that supports them.

\begin{table}[t]
\caption{High-level application areas in the corpus.}
\label{tab:domains}
\centering
\footnotesize
\setlength{\tabcolsep}{3.0pt}
\begin{tabularx}{\columnwidth}{X r X}
\toprule
Application area & Repos. & Representative repositories \\
\midrule
Numerical computing & 12 & LAPACK, PETSc, QUADPACK \\
Earth and engineering & 11 & WPS, Nek5000, OpenFAST \\
Data and utilities & 10 & HDF5, JSON-Fortran, M\_time \\
Chemistry and materials & 6 & CP2K, xTB, Packmol \\
Language and developer tools & 6 & stdlib, fpm, test-drive \\
ML and data analysis & 5 & neural-fortran, SISSO, PyMC2 \\
\bottomrule
\end{tabularx}
\end{table}

We assigned repositories to three ordered structural-complexity tiers--low, medium, and complex--using source size, inter-module coupling, and advanced Fortran features. Table~\ref{tab:composition} reports the resulting composition; Fig.~\ref{fig:overview} shows why size alone does not determine a tier.

\begin{table}[t]
\caption{Corpus composition measured from Git-tracked source at each pinned revision.}
\label{tab:composition}
\centering
\setlength{\tabcolsep}{3.2pt}
\begin{tabular}{lrrrr}
\toprule
Tier & Repos. & Files & Source lines & Probes \\
\midrule
Low     & 20 &   768 &   233,077 & 203 \\
Medium  & 20 & 2,313 &   916,989 & 201 \\
Complex & 10 & 9,035 & 5,419,740 & 157 \\
\midrule
Total   & 50 & 12,116 & 6,569,806 & 561 \\
\bottomrule
\end{tabular}
\end{table}

\subsection{Descriptive profile}

Repository size varies by more than three orders of magnitude. The median repository has 20,179.5 Fortran source lines, while individual pins range from 1,448 to 1,571,444 lines and from three to 3,600 tracked Fortran files. Within tiers, median size rises from 5,895 lines (low) to 27,186 (medium) and 318,107 (complex). The overlap visible in Fig.~\ref{fig:overview} reflects the additional use of coupling and language features during tier assignment.

The corpus spans reusable libraries, developer infrastructure, and full simulation systems across the areas in Table~\ref{tab:domains}. This breadth tests migration methods on numerical kernels, public APIs, build recovery, cross-module state, data formats, and application-scale coupling. Repository- and tier-level results can therefore reveal differences hidden by a single corpus-wide score.

\subsection{Immutable acquisition and measurement}

The acquisition script fetches each recorded Git object and checks it out in a detached-HEAD state; moving default branches never serve as benchmark identifiers. File and line totals are recomputed from paths returned by \texttt{git ls-files}. We define a Fortran source file as one with a case-insensitive suffix in \{\texttt{.f}, \texttt{.for}, \texttt{.f90}, \texttt{.f95}, \texttt{.f03}, \texttt{.f08}\}. The line count includes comments and blank lines. Counting only Git-tracked files at the pinned revision excludes generated build artefacts and local changes, making the totals reproducible.

To respect third-party licences, \fx records repository origins and revision identifiers but does not vendor the source corpus. Forty-seven pins have an unambiguous project-level licence expression. One has a mixed or ambiguous licence, and two have no explicit licence at the pinned revision; all three are marked \texttt{NOASSERTION}. Users must consult the recorded evidence before redistributing acquired source. This revision-aware approach addresses known hazards in mining GitHub~\cite{kalliamvakou2016perils}.

\subsection{Behavioural probe curation}

The catalogue fixes the inputs and comparison rules for 561 checks. We call each execution with a fixed input and comparison rule a \emph{probe}. A \emph{pinned routine} is an upstream Fortran function or subroutine from the exact Git revision in the repository manifest. A \emph{direct probe} compiles an original pinned file with a small driver. This driver is a complete, minimal Fortran program that supplies the fixed input, calls the named routine, and prints its output. It contains no alternative implementation of the behaviour. For example, a Cmathtuts driver assigns 3 and 4, calls the pinned \texttt{pythagoras} routine, and prints 5; it does not compute the square root itself. The harness creates the driver in a temporary build directory, compiles it with the pinned file, runs it, and then discards the generated source and executable. The versioned harness code records how to recreate the driver. Multiple probes may use different inputs for the same routine. Probe counts therefore measure tested executions, whereas distinct repository--routine pairs measure functional breadth. For example, the $(3,4)$ and $(-5,12)$ cases are two direct probes of one pinned routine.

Each probe is linked to its executable definition and source information. Of the 561 probes, 446 use an original pinned file plus driver, 8 are text-matched extractions, 73 are inline reimplementations, and 34 are synthetic fixtures. Before any target was run, an outcome-neutral curation protocol required direct evidence in every repository, at least five direct probes per repository, and execution of at least two distinct routines in every repository. The direct probes span 131 repository--routine pairs. The accompanying provenance documentation records exact upstream locations, interface-only dependency stubs, and retained disagreements.

Probe allocation balances representation across tiers: low, medium, and complex repositories contribute 36.2\%, 35.8\%, and 28.0\% of the checks. This prevents the ten complex systems, which contain 82.5\% of the source lines, from determining the aggregate result. Comparators support scalar, array, Boolean, and duration outputs. The 280 numeric cases use shape checks and case-specific absolute and relative tolerances. If a target cannot run a case, the adapter records an error instead of silently omitting it.

Future versions will prioritise additional public routines and replace the remaining weak contracts with direct execution of pinned source. Each change to the benchmark definition receives a new version, so results from a release always refer to the same set and number of probes.

\section{Representation, Tooling, and Use}

\subsection{Storage and integrity}

Table~\ref{tab:schema} summarises the versioned records and fields that define version~1.0.0. The release also includes a dataset card, methodology and schema documents, Apache-2.0 code, CC BY~4.0 metadata and documentation, a changelog, and a citation file.

\begin{table}[t]
\caption{Versioned records and their reproducibility role.}
\label{tab:schema}
\centering
\footnotesize
\setlength{\tabcolsep}{3.5pt}
\begin{tabularx}{\columnwidth}{l X}
\toprule
Record & Required information \\
\midrule
Repository & Stable ID, harness name, checkout directory, original GitHub URL, 40-character SHA, revision date, tier, domain, file/line counts, SPDX expression, licence status and evidence \\
Probe & Repository and probe IDs, case name, comparison mode, and executable definition path \\
Source information & Probe ID, evidence category, original repository file, routine and line range when applicable, and executable definition \\
Schema & JSON type and required-field constraints for all CSV records \\
Definition & Fixed input, source invocation, normalisation, tolerance, and target resolution \\
\bottomrule
\end{tabularx}
\end{table}

The standard-library validator checks manifest structure, identifiers and cross-references, expected totals, required files and source locations, repository revisions and URLs, probe allocation, and licence declarations. Versioning rules prohibit silent changes to a pin, input, or tolerance after release. Corrections increase the benchmark version and are recorded in the changelog. Users can therefore inspect the release without first acquiring several gigabytes of original source:
\begin{quote}\small
\texttt{python scripts/validate.py}\\
\texttt{python scripts/summarize.py}
\end{quote}
One exact pin can then be acquired with \texttt{bash clone\_corpus.sh --repo roots-fortran}; \texttt{--check-metrics} recomputes its numbers of Git-tracked Fortran files and lines. Omitting \texttt{--repo} acquires all 50 pinned revisions.

\subsection{Adapters and reporting contract}

The bundled adapter compiles the Fortran reference programs with \texttt{gfortran}, invokes the Python implementation under evaluation, normalises only what each probe permits, and applies the shared comparator. An adapter for another target replaces only the target-resolution and invocation logic. It must retain the probe inputs, reference-program execution, comparison modes, and tolerances. It must also declare name mappings before inspecting outcomes and report all non-runnable cases. We recommend reporting agreements out of all 561 planned cases, together with per-repository and per-tier results. Studies may also report agreement among runnable cases, but must label this as a separate executable-only result and state how many cases it excludes.

A clean-room replay remeasures an exact pin, compiles seven legacy wrappers, and runs one check end to end. On macOS~26.7 ARM64 with CPython~3.13.0 and GNU Fortran~15.2.0, the acquired \texttt{roots-fortran} revision matched its recorded size, and probe \texttt{f2x50-low-roots-fortran-001} passed. The same smoke test also passed in the supplied Debian Bookworm container on a Linux host. An accounting replay against archived Python targets from the earlier migration study reported all planned cases: 436 agreements, 74 mismatches, and 51 errors, totalling 561. This tested the adapter's denominator and failure reporting, not a competing migration method; the targets are not part of the benchmark release. The predeclared distinct-routine selection retained every selected case, including both \texttt{Cmathtuts:pythagoras} inputs. The release includes pinned dependencies, a cross-platform running guide, and a Dockerfile; its metadata distinguishes build, missing-target, exception, shape, and value failures.

\subsection{Worked probe example}

The seven \texttt{roots-fortran} probes illustrate the workflow. Acquisition checks out the recorded revision and can recompute its manifest counts. Each check compiles \texttt{src/root\_module.F90} from the pinned revision with a small driver. Four bisection cases evaluate $f(x)=x^2-2$ on a standard positive bracket, a wider positive bracket, a negative bracket, and an interval with no sign change. Three further cases invoke the Brent, Ridders, and Illinois solvers on $[1,2]$. For each fixed input, the source executable emits the root, residual, and status flag. In a benchmark study, the adapter invokes the corresponding target calls with the same function and brackets, then passes their outputs to the shared numeric comparator.

The bisection observations use absolute and relative tolerances of $10^{-10}$. The three method probes use an absolute tolerance of $10^{-6}$ to allow documented differences in solver stopping rules. If a target does not expose one method, that probe is recorded as non-runnable and still counts among the 561 planned cases. The source-information manifest identifies all seven probes as original-file-plus-driver cases. It also records the original repository file, routine, line range, and executable definition. A consumer can restrict an analysis to direct-source cases or report each category separately.

For the clean-room demonstration, the release includes a small Python implementation of bisection and a script that compares it with the pinned Fortran routine. The script runs \texttt{f2x50-low-roots-fortran-001}, one of the 561 published checks, and produces a passing reference result. It demonstrates the complete evaluation path without claiming that the entire \texttt{roots-fortran} repository has been migrated. To reproduce it, a user validates the manifests, runs \texttt{clone\_corpus.sh --repo roots-fortran --check-metrics}, and invokes the reference example. The repository documentation lists the compiler and NumPy prerequisites. Intermediate executables are created in temporary directories rather than committed to the dataset.

\subsection{Research uses and FAIR properties}

\fx supports four immediate research questions: (RQ1) How do repository size, coupling, build structure, and Fortran features affect migration? (RQ2) Which failure modes persist across targets or strategies? (RQ3) Do translators preserve numerically sensitive behaviour? (RQ4) How do build reconstruction and API mapping affect the number of executable probes?

These questions support three study designs. A single-system evaluation can analyse all 561 outcomes by tier, repository, comparison mode, and failure category. A paired comparison can apply two methods or targets to the same pins and probes. A longitudinal study can run newer tools against unchanged version~1.0.0 inputs while publishing adapters and results separately. Repository-level summaries should accompany the micro-averaged score so that similar checks in one project do not dominate.

The benchmark should not serve as both training data and unseen evaluation data. Studies that tune prompts, mappings, or repair rules on \fx should disclose this use, separate development repositories before inspection, and avoid describing results on the same release as out-of-sample performance. Contract agreement measures preservation only for the published inputs, and its evidential strength varies across source-evidence categories. Broader migration claims should also consider build success, static quality, human review, performance, or expanded tests.

The release follows the FAIR principles~\cite{wilkinson2016fair}. Stable IDs, citation metadata, and a versioned archive make it findable. HTTPS acquisition and open manifests make it accessible. CSV, JSON Schema, Git SHAs, and SPDX expressions support interoperability. Explicit licences, source traceability, executable definitions, validation, and immutable versioning support reuse. Version~1.0.0, including setup instructions and one-command validation, is available at \url{https://github.com/ShraddhaSurana/f2x50}. The archived release is available at \href{https://doi.org/10.5281/zenodo.22890055}{doi:10.5281/zenodo.22890055}.

\subsection{Evolution and maintenance protocol}

Future versions may add repositories, probes, or target adapters, but these changes have different meanings. A target adapter is an external consumer and does not change the benchmark's identity. A new probe or source pin changes the set or total number of evaluation cases and therefore requires a new release. Proposed probes must be deterministic, publicly describable, runnable on modest hardware, selected without inspecting a method's outcome, and assigned to one of the published comparison families. Pull requests are checked for schema validity, unique stable IDs, manifest joins, executable definitions, and aggregate totals. Released manifests remain immutable; the changelog and archival metadata link errata and superseding versions. This policy allows later coverage improvements while preserving exact reproduction of results reported against version~1.0.0.

\section{Limitations, Future Work, and Conclusion}

The corpus is a purposive sample of Fortran software, and the expert-assigned tiers reflect the stated size, coupling, and language criteria. Source-line counts include comments and blank lines. Some drivers still contain interface-only stubs, and the three \texttt{NOASSERTION} pins require separate permission analysis before redistribution.

Future work will prioritise direct execution for the remaining inline contracts, coverage-guided expansion to unrepresented routines, richer static descriptors, and more target adapters. These changes will be versioned without rewriting version~1.0.0.

\fx provides a revision-pinned Fortran corpus and reusable migration infrastructure. Its traceable boundary between source evidence and migration outputs supports comparable studies across tools and target ecosystems.

\IEEEtriggeratref{6}
\bibliographystyle{IEEEtran}
\bibliography{references}

\end{document}